\documentclass[preprint]{aastex}

\newcommand{\be}{\begin{equation}}
\newcommand{\ee}{\end{equation}}
\newcommand{\pprime}{{\prime\prime}}

\begin{document}

\title{ON THE EXISTENCE OF THREE-DIMENSIONAL \\
HYDROSTATIC AND MAGNETOSTATIC EQUILIBRIA \\
OF SELF-GRAVITATING FLUID BODIES}

\author{Daniele Galli}
\affil{INAF-Osservatorio Astrofisico di Arcetri \\
              Largo Enrico Fermi 5 \\
              I-50125 Firenze, Italy \\
              galli@arcetri.astro.it}


\begin{abstract}
We develop an analytical spectral method to solve the equations of
equilibrium for a self-gravitating, magnetized fluid body, under the
only hypotheses that ({\it a}\/) the equation of state is isothermal,
({\it b}\/) the configuration is scale-free, and ({\it c}\/) the body
is electrically neutral. All physical variables are represented as
series of scalar and vector spherical harmonics of degree $l$ and order
$m$, and the equilibrium equations are reduced to a set of coupled
quadratic algebraic equations for the expansion coefficients of the
density and the magnetic vector potential. The method is general,
and allows to recover previously known hydrostatic and magnetostatic
solutions possessing axial symmetry.  A linear perturbation analysis of the
equations in spectral form show that these basic axisymmetric states,
considered as a continuos sequence with the relative amount of magnetic
support as control parameter, have in general no neighboring
nonaxisymmetric equilibria. This result lends credence to a conjecture
originally made by H. Grad and extends early results obtained by E.
Parker to the case of self-gravitating magnetized bodies.  The only
allowed bifurcations of this sequence of axisymmetric equilibria are represented by
distortions with dipole-like angular dependence ($l=1$) that can be
continued into the nonlinear regime.  These new configurations are
either ({\it i}\/) azimuthally asymmetric ($m=\pm 1$) or ({\it
ii}\/) azimuthally symmetric but without reflection symmetry with
respect to the equatorial plane ($m=0$).  It is likely that these
configurations are not physically acceptable solutions for isolated
systems, but represent instead the manifestation of a general {\it
gauge freedom} of self-similar isothermal systems.  To the extent that
interstellar clouds can be represented as isolated magnetostatic
equilibria, the results of this study suggest that the observed triaxial shapes of
molecular cloud cores can be interpreted in terms of weakly damped 
Alfv\`en oscillations about an equilibrium state.
\end{abstract}

\keywords{Hydrodynamics, Magnetohydrodynamics, Molecular Clouds}

\section{Introduction}

Much of what we know of magnetostatic (MS) equilibria in plasma physics
and astrophysics has come from studying systems with one ignorable
coordinate: axisymmetric systems are one example. The symmetry associated
to with one ignorable coordinate conveniently allows MS problems to
be reduced to neat tractable forms. In the three-dimensional case,
no one has ever been able to find nonaxisymmetric solutions of the MS
equations for a self-gravitating fluid, and their actual existence has
remained dubious (see Sect.~1.2). The goal of this paper is to explore 
the the existence of configurations of equilibrium of self-gravitating fluid
bodies in the presence of a large-scale magnetic field without symmetry
restrictions,  with an application in mind to the study of the densest 
parts of molecular clouds (the so-called cloud cores), the sites of star formation.

The motivation for this work is the following. The dense cores of
interstellar molecular clouds are observed to be close to a state of
virial equilibrium (balance of gravitational, thermal and magnetic
energy (see e.g. Myers~1999), yet their shapes are distinctly
non-spherical: the best fit to the projected axial ratio distribution
is obtained with triaxial ellipsoids (Jones \& Basu~2002; Goodwin,
Ward-Thompson, \& Whitworth~2003). The question is then to consider the
possible role of the anisotropic forces associated to the magnetic
fields present in these cores (see e.g. Crutcher~2001) to
allow the existence of shapes of non-trivial topology.

The paper is organized as follows: in Sect.~2 we summarize the
available results on the equilibrium of ideal (i.e. non viscous and non
resistive) plasmas, discussing in particular the case of
non-selfgravitating configurations; in Sect.~3 we formulate
mathematically the problem and introduce non-dimensional
self-similar variables; in Sect.~4 we describe the method of solution,
based on the expansion of the angular part of the variables in scalar
and vector spherical harmonics; in Sect.~5 and 6 we show that our method of
solution recovers previously known axially symmetric hydro- and
magnetostatic equilibria, respectively; in Sect.~7 we linearise the
equations around these basic equilibria, and look for neutral modes
(neighbouring equilibria) possessing a different symmetry than the
basic states; finally, in Sect.~8 we summarize our results and discuss
their implications for the observed shapes of molecular
cloud cores.

\section{Ideal magnetostatic equilibria}

Consider a magnetized, isothermal, self-gravitating fluid body satisfying
the ideal MS equations
\be
c_{\rm s}^2\nabla\rho+\rho\nabla {\cal V}=
{1\over 4\pi}(\nabla\times {\bf B})\times {\bf B},
\label{eqforce}
\ee
\be
\nabla\cdot{\bf B}=0,
\label{divB}
\ee
\be
\nabla^2 {\cal V}=4\pi G\rho,
\label{eqPoisson}
\ee
where ${\bf B}$ is the magnetic field, $\rho$ is the density, ${\cal V}$
is the gravitational potential, $G$ is the constant of gravitation and
$c_{\rm s}$ is the sound speed.  
With applications to interstellar molecular clouds in mind,
we have assumed here an isothermal equation
of state. All known solutions of the 
set of equations (\ref{eqforce})--(\ref{eqPoisson})
are characterized by a coordinate symmetry that reduces the
number of variables from three to two or one. The symmetry associated
with an ignorable coordinate allows MS problems to be reduced to a
second-order elliptic partial differential equation (Dungey~1953),
conventionally called the {\it Grad-Shafranov equilibrium equation}. These
symmetric equilibria may be grouped into ({\it i}\/) {\it axisymmetric}
($\partial/\partial\varphi=0$), ({\it ii}\/) {\it cylindrically symmetric}
($\partial/\partial z=0$), and ({\it iii}\/) {\it helically symmetric}
systems ($\partial/\partial\varphi=k\,\partial/\partial z$).  As shown
by Edenstrasser (1980), helical symmetry represents the most general
admissible invariance property of the MS equations, with rotational and
translational invariance as limiting cases.

Applications of eq.~(\ref{eqforce})--(\ref{eqPoisson}) to the study of
interstellar molecular clouds have largeley focused on axisymmetric
magnetic configurations (e.g. Mouschovias 1976; Nakano 1979; Mestel \&
Ray~1985; Tomisaka, Ikeuchi,\& Nakamura~1988; Barker \& Mestel~1990).
Cylindrically symmetric equilibria have also been extensively studied,
originally in connection with the stability of galactic spiral arms
(Chandrasekhar \& Fermi 1953; Stod{\'o}{\l}kiewicz 1963), and later as
models of filamentary clouds (Nagasawa~1987; Nakamura, Hanawa, \&
Nakano 1995).  In the latter context, magnetic configurations
possessing helical symmetry have also been explored (Nakamura, Hanawa,
\& Nakano~1993, Fiege \& Pudritz 2000a,b). An excellent introduction 
to the subject can be found in the monography by Mestel~(1999).

\subsection{Parker's theorem}

For non-selfgravitating plasmas, the fundamental question as to the
existence of MS equilibria of a more general symmetry, or with no
symmetry at all (3-D MS equilibria), has been formulated several
times (e.g. Low 1980, Degtyarev et al.~1985) but never properly
answered. Grad~(1967) conjectured that, with ${\cal V}=0$, only ``highly
symmetric'' solutions of the system (\ref{eqforce})--(\ref{divB}) should
be expected, in order to balance the highly anisotropic Lorentz force with
with pressure gradients and gravity, which are forces involving scalar
potentials.  In a fundamental paper, Parker (1972) proved rigorously the
non existence of 3-D MS equilibria that are small perturbations of 2-D
equilibria having translational symmetry. This result is conventionally
referred to as {\it Parker's theorem}.

On the basis of this result, Parker~(1979) argued that realistic
magnetic fields with no well defined symmetries must evolve in a
genuinely time-dependent way, until all non-symmetric components of the
field are destroyed by dissipation and reconnection and the topology
becomes symmetric, a process known as {\it topological non-equilibrium}
of configurations lacking high degrees of symmetry (see also Tsinganos,
Distler, \& Rosner~1984).  The numerical simulations of Vainshtein et
al.~(2000) provide a striking illustration of this process.  The
problem with interstellar clouds is that ohmic dissipation times are of
the order of $\sim 10^{15}$~yr, and therefore the kind of monotonic
relaxation to equilibrium envisaged by Parker~(1979) can be ruled out.
For a non-dissipative plasma, Moffatt~(1985, 1986) has shown
that {\it stable} non-axisymmetric magnetostatic equilibria of
non-trivial topology do exist, but may require the presence of
tangential field discontinuities (current sheets). The
mathematical method adopted in this paper, based on analytical functions
and regular perturbation expansions, cannot address the question of the existence
of Moffatt-type equilibria in self-gravitating ideal plasmas. However, given
the general character of Moffatt's conclusions, the question is 
considered again at the light of our results in Sect.~8.

Does the presence of fluid motions modify this picture?
Tsinganos~(1982) found that Parker's theorem remains valid for steady
dynamical (${\bf v}\neq 0$) configurations possessing translational
invariance, and stressed the analogy between this result and the familiar
Taylor-Proudman theorem of hydrodynamics. However, Galli et al. (2001)
found that a class of two-dimensional axisymmetric MHD equilibria
(rotating, magnetized, self-gravitating singular isothermal disks)
does have neighboring non-axisymmetric equilibrium states, provided the
rotation speed becomes sufficiently high (supermagnetosonic). Thus,
not only the presence of fluid motions, but also the geometry of the
system, seems to play a crucial role in breaking the symmetry of the
equilibrium state.

As for the effect of gravity, Field (1982, quoted by Tsinganos et
al.~1984) objected that the neglect of the constraining effect exerted by
the plasma's self-gravity may severely limit the domain of existence of
MS equilibria. In a series of papers (Low 1985; Bogdan \& Low 1986; Low
1991), Low and collaborators have elaborated a general method to solve
the equations of MS in the presence of an external gravitational field,
but the problem presents considerable mathematical difficulties. In
agreement with Grad's conjecture, Low concluded that applying some form
of symmetry to the magnetic field is probably essential for the existence
of equilibrium, in order to balance the highly anisotropic Lorentz force
with pressure gradients and gravity, which are forces involving scalar
potentials.  In the special case of an imposed gravitational field,
either uniform or due to a point mass, Low was able to find families of
three-dimensional MS solutions.  For a self-gravitating gas, Low~(1991)
showed that the problem can be reduced to the solution of two coupled
partial differential equations for two unknown functions, but did not
proceed further.

\section{The equations of the problem}

In the following we specialize to scale-free equilibria.  Working in
spherical coordinates $(r,\theta,\varphi)$, we assume that every physical
quantity of the problem can be factorized in one function (power-law or
logarithmic) of $r$ times a function of $\theta$ and $\varphi$ only. This 
assumption allows a considerable simplification of the equations of the 
problem. We then show that the magnetic field must be {\it poloidal}
in the Stratton-Chandrasekhar classification (Sect.~3.1), and we derive
the governing set of non-dimensional equations (Sect.~3.2). Finally, we
obtain the equations of the problem in the special case of axial symmetry
(Sect.~3.3).

The assumption of a power-law (or logarithmic) dependence on radius for
the variables of the problem can be justified on observational and
theoretical grounds.  The observed density profiles of cloud cores
approach in general a $r^{-2}$ power-law behaviour in the outer parts,
but are flatter near the cloud's centre (see e.g. Ward-Thompson, Motte,
\& Andr\'e~1999). Idealised models that assume a power-law behaviour of
the density (and the intensity of the magnetic field, etc.) over all
radii are justified by our expectation that the evolution of the
observed cloud cores tends to a singular configuration. The driving
process responsible for this evolution has been identified in the
ambipolar diffusion of a weakly ionised gas, as originally proposed by
Mestel \& Spitzer (1956). Several numerical calculations of cloud
evolution driven by ambipolar diffusion show that the density profile
steepens to $r^{-2}$ to arbitrary number of decades in radius as the
singular state is approached (Fiedler \& Mouschovias~1993; Ciolek \&
Mouschovias~1993, 1994; Basu \& Mouschovias~1994, 1995). Thus,
scale-free (singular) configurations provide realistic models for
molecular cloud cores in the so-called ``pivotal'' state, i.e. on the
verge of gravitational collapse.  The price to pay for this simplifying
assumption, is, of course, the introduction of an artificial
singularity at the origin, where the density and the magnetic field
diverge.

\subsection{Toroidal and poloidal fields}

Any solenoidal vector field {\bf B} can be expressed as a linear
combination of a certain basic {\it toroidal} (${\bf B}_{\rm t}$) and {\it
poloidal} (${\bf B}_{\rm p}$) fields (Stratton~1941, Chandrasekhar~1961),
given by
\be
{\bf T}=\nabla\left({\Theta\over r}\right)\times {\bf r},
\ee
and
\be
{\bf S}=\nabla\times\left[\nabla\left({\Psi\over r}\right)\times {\bf r}\right],
\ee
where $\Theta$ and $\Psi$ are arbitrary scalar functions of position.
In addition to have zero divergence, the toroidal and poloidal fields
${\bf T}$ and ${\bf S}$ are characterized by vanishing radial component
($T_r=0$) and vanishing radial component of the curl ($[\nabla\times
{\bf S}]_r=0$), respectively\footnote{Notice that this terminology
is different from the one commonly adopted in astrophysics, where is
customary to call {\it poloidal} a field with components only along $r$
and $\theta$, and {\it toroidal} a field with only a component along
$\varphi$.}.

For a self-similar problem, if there is no radial current at one $r$, then the
same is true for all $r$.  We want it to be true at large $r$ because
otherwise there would be a flow of charge to infinity, and the cloud
would become electrically charged. Thus, in our problem, the curl of
the magnetic field has zero radial component, and is therefore
{\it poloidal} in the Stratton-Chandrasekhar
terminology, 
\be
{\bf B}={\bf S}
=\nabla\times\left[\nabla\left({\Psi\over r}\right)\times {\bf r}\right].
\label{defB}
\ee
In axial symmetry, we recover the condition 
${\bf B}\cdot \nabla \Phi = 0$, where $\Phi
\equiv -2\pi \sin\theta \partial \Psi/\partial \theta$ is the usual 
magnetic flux function. 

\subsection{Nondimensional self-similar variables}

It is easy to see from eq.~(\ref{eqforce})--({\ref{eqPoisson}) and
eq.~(\ref{defB}) that if $\rho$, ${\cal V}$ and $\Psi$ are chosen to
have appropriate power-law dependences in $r$ one can eliminate all
$r$-dependences in the MS equations for an isothermal gas.  Following Li
\& Shu~(1996, hereafter LS96) we adopt the scaling
\be
\rho={c_{\rm s}^2\over 2\pi Gr^2} R(\theta,\varphi),
\label{scalrho}
\ee
\be
{\cal V}=2c_{\rm s}^2[(1+H_0)\ln r+V(\theta,\varphi)],
\label{calV}
\ee
\be
\Psi={2 c_{\rm s}^2r\over \sqrt{G}}F(\theta,\varphi),
\label{scalPsi}
\ee
where the quantity $H_0$ in the expression of the gravitational potential
eq.~(\ref{calV}) is a dimensionless constant to be specified.  We also
define the angular parts of the $\nabla$ and $\nabla^2$
operators,
\be
\nabla_\Omega\equiv {\bf\hat\theta}{\partial\over\partial\theta}+
{{\bf\hat\varphi}\over\sin\theta}{\partial\over\partial\varphi},
~~~~~~\nabla^2_\Omega \equiv {1\over \sin \theta}{\partial\over \partial \theta}
\sin\theta {\partial \over \partial \theta} + {1\over \sin^2\theta}
{\partial^2\over \partial \varphi^2}.
\ee
Inserting eq.~(\ref{scalPsi}) into eq.~(\ref{defB}) we obtain
the expression for the magnetic field 
\be
{\bf B}={2 c_{\rm s}^2\over \sqrt{G} r}[A{\bf\hat r}+\nabla_\Omega F],
\label{bfield}
\ee
where we have defined
\be
A\equiv -\nabla^2_\Omega F.
\label{defA}
\ee
Finally, the Lorentz force is given by 
\be 
{1\over
4\pi}(\nabla\times {\bf B})\times {\bf B}={c_{\rm s}^4\over \pi G r^3}
[(\nabla_\Omega F\cdot\nabla_\Omega A){\bf\hat r}- A\nabla_\Omega A].  
\ee 
{\bf Notice that a magnetic field with no radial component of its curl
and proportional to $r^{-1}$, as assumed here, cannot be force-free.}

Inserting expressions (\ref{scalrho})--(\ref{scalPsi}) into
eq.~(\ref{eqforce})--(\ref{eqPoisson}), we rewrite the condition of 
force balance in the radial direction as
\be 
H_0 R= \nabla_\Omega F\cdot \nabla_\Omega A,
\label{radfor} 
\ee 
and the condition of force balance in the tangential direction as
\be 
{1\over 2}\nabla_\Omega R + R\nabla_\Omega
V+A\nabla_\Omega A=0. 
\label{tanfor} 
\ee 
Finally, Poisson's equation becomes
\be 
\nabla^2_\Omega V=R-(1+H_0). 
\label{pois}
\ee 
We now show that eq.~(\ref{radfor})--(\ref{pois}) generalize to the set of
equations derived by LS96 for axisymmetric isothermal scale-free equilibria.

\subsection{The axisymmetric case}

Assuming $\partial /\partial \varphi=0$, and defining 
$\phi(\theta)\equiv -\sin\theta F^\prime$, we obtain 
\be
A(\theta)={\phi^\prime \over \sin\theta},
\ee
where a prime indicate derivation with respect to $\theta$.
With this definition, eq.~(\ref{radfor}) reduces to eq.~(13) 
of LS96,
\be
{d\over d\theta}\left({\phi^\prime\over \sin\theta}\right)=-H_0{R\over\phi}\sin\theta.
\label{ls13}
\ee
Poisson's eq.~(\ref{pois}) reads
\be
{1\over\sin\theta}{d\over d\theta}\left(\sin\theta{dV\over d\theta}\right)=R-1-H_0.
\ee
Eliminating $dV/d\theta$ using eq.~(\ref{tanfor}), and simplifying the result
with the help of eq.~(\ref{ls13}), we obtain eq.~(12) of LS96,
\be
{1\over\sin\theta}{d\over d\theta}\left[\sin\theta\left(2H_0{\phi^\prime\over\phi}-
{R^\prime\over R}\right)\right]=2(R-1-H_0),
\label{ls}
\ee
that completes the set of equations governing axisymmetric, scale-free,
isothermal equilibria.

Pure hydrostatic equilibria are described by the single equation
\be
{1\over\sin\theta}{d\over d\theta}\left(\sin\theta{R^\prime\over R}\right)
=2(1-R),
\ee
obtained by setting $\phi=0$ and $H_0=0$ in eq.~(\ref{ls}).
Medvedev \& Narayan~(2000, hereafter MN00) have found an analytical 
solution of this equation, 
\be
R(\theta)={1-e^2\over (1\pm e\cos\theta)^2},
\label{mn}
\ee
where iso-density contours (described by $r(\theta)\propto
\sqrt{R(\theta)}$ according to eq.~[\ref{scalrho}]) are {\it confocal 
ellipsoids} of eccentricity $e$, with $0<e<1$. For $e=0$, this solution
reduces trivially to the singular isothermal sphere, with $R=1$. 
We will return to this family of equilibria in Sect.~5.

\section{Method of solution}

At first sight, the problem appears to be overconstrained, since
there are four unknown functions ($R$, $V$, $A$ and $F$) and 
four equations (eq.~[\ref{defA}], [\ref{radfor}], [\ref{tanfor}] and 
[\ref{pois}]), one of which (eq.~[\ref{tanfor}]) has two components.
However, it is easy to see that the vectors $\nabla_\Omega R$,
$\nabla_\Omega V$ and $\nabla_\Omega A$ in eq.~(\ref{tanfor}) are {\it
parallel}, thus the condition of force balance in the tangential
direction reduces to one constraint only. To
see this, first take the curl of eq.~(\ref{tanfor}), obtaining
$\nabla_\Omega R\times \nabla_\Omega V=0$, a condition implying that
isodensity and equipotential surfaces are coincident.
Then take the vector product
of eq.~(\ref{tanfor}) with $\nabla_\Omega R$, obtaining $\nabla_\Omega
R\times \nabla_\Omega A=0$. It follows then that $\nabla_\Omega V\times
\nabla_\Omega A=0$ (CVD). 
The meaning of the condition $\nabla_\Omega V\times \nabla_\Omega
A=0$ can be understood by writing the expression for the electric
current ${\bf j}$ from Amp\`ere's law using eq.~(\ref{bfield}),
\be
{\bf j}={c\over 4\pi}\nabla\times {\bf B}
=-{c_{\rm s}^2 c\over 2\pi \sqrt{G} r^2}{\bf\hat r}\times \nabla_\Omega A.
\label{curr}
\ee
Eq.~(\ref{curr}) shows that the electric current ${\bf j}$ is 
perpendicular to $\nabla_\Omega A$, and therefore implies that 
the current flows over equipotential (or isodensity) surfaces. 

The method adopted in this work is based on the expansion of the vector
variables of the problem in vector spherical harmonics.  Any vector
quantity is represented by a convergent infinite sequence of complex
coefficients, and the problem is formulated in the Hilbert space.
The solution of the nonlinear partial differential equations of the
problem is thus reduced to the solution of infinite set of nonlinear
algebraic equations. The coupling coefficients are expressed in terms of
Wigner $3j$ symbols, and are evaluated by the Racah's formula (see 
Appendix A). 

\subsection{Multipole expansion}

Vector spherical harmonics (see e.g. Arfken~1985) provide 
the natural basis for a multipole
expansion of eq.~(\ref{radfor})--(\ref{pois}). Here we follow Morse \&
Feshbach~(1953) defining
\be
{\bf P}_{lm}=Y_{lm}{\bf\hat r},~~~~{\bf B}_{lm}=
{1\over\sqrt{l(l+1)}}\nabla_\Omega Y_{lm},~~~~{\bf C}_{lm}=
-{\bf\hat r}\times {\bf B}_{lm}.
\label{vectharm}
\ee
As we will see, the properties of the product of vector spherical
harmonics allow non-linear terms to be dealt with in a systematic way
by Wigner $3j$ symbols (or Clebsch-Gordan coefficients).

Clearly, an infinite number of orthonormal bases can be generated from the
${\bf P}_{lm}$, ${\bf B}_{lm}$, ${\bf C}_{lm}$ basis by the application of
unitary transformation. The set (\ref{vectharm}) is especially convenient
for situations where a preferred radial direction is present, like in our
case, since the ${\bf P}_{lm}$ are {\it radial} whereas the ${\bf B}_{lm}$
and ${\bf C}_{lm}$ are {\it tangential} to the unit sphere.  
Thus, if we expand $R$, $V$, $A$ and $F$ in spherical harmonics, and we
eliminate $F$ and $V$ by using eq.~(\ref{defA}) and (\ref{pois}), the
remaining equations of the problems for $R$ and $A$ are {\it naturally}
expressed in the basis (\ref{vectharm}), with $\nabla_\Omega A$ and
$\nabla_\Omega R$ expanded in series of ${\bf B}_{lm}$ and ${\bf j}$
in series of ${\bf C}_{lm}$.  For the reasons explained in Sect.~4, we
expect therefore the expansion along ${\bf C}_{lm}$ of the equation of
force balance (eq.~[\ref{eqforce}]) to be trivially satisfied.  

We expand the functions $R$, $V$, $A$ and $F$ in spherical harmonics,
\be
R(\theta,\varphi)=1+H_0+\sum_{lm}R_{lm}Y_{lm}(\theta,\varphi),
\label{Rexp}
\ee
\be
V(\theta,\varphi)=\sum_{lm}V_{lm}Y_{lm}(\theta,\varphi),
\label{Vexp}
\ee
\be
A(\theta,\varphi)=\sum_{lm}A_{lm}Y_{lm}(\theta,\varphi),
\label{Aexp}
\ee
\be
F(\theta,\varphi)=\sum_{lm}F_{lm}Y_{lm}(\theta,\varphi),
\label{Fexp}
\ee
where the sum is for $l\ge 1$ and $-l\le m \le l$. In general,
$R_{lm}$, $V_{lm}$, $A_{lm}$ and $F_{lm}$ are complex coefficients.
Since $R$, $V$, $A$ and $F$ are real functions, we have to require that
\be
R_{l-m}=(-1)^m R^\ast_{lm},~~~\mbox{etc.}
\label{realcoe}
\ee
The constant factor $1+H_0$ in the expansion of $R(\theta,\varphi)$
is chosen to simplify Poisson'e equation~(\ref{pois}). Notice that 
with this choice
\be
\oint R\;d\Omega=4\pi(1+H_0),
\ee
a condition equivalent to the ``integral constraint'' of LS96 (their
eq.~[18]). Therefore, as in LS96, $H_0$ measures the fractional increase
in the mean density that arises because the magnetic field contributes
to support the cloud against self-gravity.

It is straightforward to compute the spherical mass-to-flux ratio,
or the ratio of the mass contained in a sphere centered on the origin
to the magnetic flux through a circle of the same radius in the 
equatorial plane.  The mass enclosed in a sphere of radius $r$ is 
\be
M={c_{\rm s}^2 r\over 2\pi G}\oint R\;d\Omega={2c_{\rm s}^2 r\over G}(1+H_0).
\ee
The magnetic flux through a circle of the radius $r$ in the 
equatorial plane is equal by Gauss theorem to the magnetic flux 
through a semisphere of radius $r$,
\be
\Phi={2c_{\rm s}^2\over \sqrt{G}r}\int {\bf B}\cdot {\bf\hat r}\;dS=
{2c_{\rm s}^2 r\over \sqrt{G}}\oint d\varphi\int_0^{\pi/2} A(\theta,\varphi)
\sin\theta\;d\theta,
\ee
where we have used the expression of ${\bf B}$ given by 
eq.~(\ref{bfield}). With the expansion (\ref{Aexp}) we obtain
\be
\Phi={4\pi c_{\rm s}^2 r\over \sqrt{G}}\sum_l A_{l0}
\int_0^{\pi/2} Y_{l0}\sin\theta\; d\theta=
{\pi c_{\rm s}^2 r\over \sqrt{G}}\sum_l \phi_l A_{l0},
\ee
where
\be
\phi_l={\sqrt{l+1}\over \Gamma\left(1-{1\over 2}l\right)
\Gamma\left({3\over 2}+{1\over 2}l\right)}.
\ee
Thus, in non-dimensional units, the spherical mass-to-flux ratio
results
\be
\lambda_r\equiv 2\pi\sqrt{G}{M\over\Phi}
={4(1+H_0)\over \sum_l \phi_l A_{l0}}.
\label{lambdar}
\ee

Using known properties of the vector spherical harmonics (see e.g.
Varshalovich, Moskalev, \& Khersonskii~1988), vector 
quantities like $\nabla_\Omega R$, $\nabla_\Omega V$, etc. are 
immediately expressed as
\be
\nabla_\Omega R=\sum_{lm}\sqrt{l(l+1)}R_{lm}{\bf B}_{lm},\quad
\nabla_\Omega V=\sum_{lm}\sqrt{l(l+1)}V_{lm}{\bf B}_{lm},~\rm{etc.}
\ee
In addition, the expansion in spherical harmonics presents the
advantage that it makes possible to solve immediately Poisson's
equation (eq.~[\ref{pois}]) and the relation between $A$ and $F$
(eq.~[\ref{defA}]), as both equations involve the angular part of
the Laplacian operator. Since
\be
\nabla^2_\Omega Y_{lm}=-l(l+1)Y_{lm},
\ee
eq.~(\ref{defA}) gives the relation between the coefficients 
$F_{lm}$ and $A_{lm}$,
\be
A_{lm}=l(l+1)F_{lm},
\label{Flm}
\ee
whereas Poisson's equation (eq.~\ref{pois}) gives the 
relation between the coefficients $V_{lm}$ and $R_{lm}$,
\be
R_{lm}=-l(l+1)V_{lm}.
\label{Vlm}
\ee

Inserting the expansions eq.~(\ref{Rexp})--(\ref{Aexp}) in
eq.~(\ref{bfield}) and eq.~(\ref{curr}), and eliminating the
coefficients $F_{lm}$ using eq.~(\ref{Flm}), we obtain the expansion in
vector spherical harmonics of the magnetic field and the electric
current,
\be
{\bf B}={2c_{\rm s}^2\over \sqrt{G} r}\sum_{lm}A_{lm}\left[{\bf P}_{lm}
+{{\bf B}_{lm}\over \sqrt{l(l+1)}}\right],
\label{Bexp}
\ee
and
\be
{\bf j}={c_{\rm s}^2 c\over 2\pi \sqrt{G} r^2}\sum_{lm}\sqrt{l(l+1)}A_{lm}{\bf C}_{lm}.
\label{jcurr}
\ee
Using the vector relation 
\be
\nabla\times {\bf C}_{lm}={1\over r}\left[\sqrt{l(l+1)}{\bf P}_{lm} + {\bf B}_{lm}\right],
\ee
we immediately recognize in the expression (\ref{Bexp}) for ${\bf B}$
the curl of a vector potential ${\bf A}$ given by
\be
{\bf A}={2 c_{\rm s}^2\over \sqrt{G}}
\sum_{lm}{A_{lm}\over \sqrt{l(l+1)}}{\bf C}_{lm}.
\ee

\subsection{Spectral form of the equations}

We first consider the radial component of the equation of force balance,
eq.~(\ref{radfor}). In terms of the vector spherical harmonics
this equation can be written
\be
\sqrt{4\pi}H_0(1+H_0)Y_{00}
+H_0\sum_{l^\prime m^\prime}R_{l^\prime m^\prime}Y_{l^\prime m^\prime}=
\sum_{l^\prime m^\prime}\;\sum_{l^\pprime m^\pprime}
\left[{l^\prime(l^\prime+1)\over l^\pprime(l^\pprime+1)}\right]^{1/2}
A_{l^\prime m^\prime}A_{l^\pprime m^\pprime}{\bf B}_{l^\prime m^\prime}
\cdot {\bf B}_{l^\pprime m^\pprime},
\label{radfor1}
\ee
where we have used eq.~(\ref{Flm}) to eliminate the coefficients $F_{lm}$.
Then, we multiply each term on both sides by $Y^\ast_{lm}$ and integrate
over solid angle $d\Omega$, using the known orthonormality properties
of spherical harmonics. It is convenient to introduce the coupling
coefficients $\alpha^{m m^\prime m^\pprime}_{ll^\prime l^\pprime}$,
defined as
\be
\alpha^{m m^\prime m^\pprime}_{ll^\prime l^\pprime}\equiv 
{1\over\sqrt{l(l+1)}}\oint 
{\bf B}_{l^\prime m^\prime}\cdot {\bf B}_{l^\pprime m^\pprime}
Y^\ast_{lm}\;d\Omega,
\label{alpha}
\ee
involving the product of three spherical harmonics (or their
derivatives). The expressions for these coupling coefficients are given
in Appendix A.  We obtain from eq.~(\ref{radfor1}) the condition
\be
\sqrt{4\pi}H_0(1+H_0)\delta_{l,0}+H_0 R_{lm}=
\sqrt{l(l+1)}\sum_{l^\prime m^\prime}\;\sum_{l^\pprime m^\pprime}
\left[{l^\prime(l^\prime+1)\over l^\pprime(l^\pprime+1)}\right]^{1/2}
\alpha^{m m^\prime m^\pprime}_{l l^\prime l^\pprime}
A_{l^\prime m^\prime}A_{l^\pprime m^\pprime}.
\label{radfor2}
\ee
The value of $\sqrt{l(l+1)}\alpha^{m m^\prime m^\pprime}_{l l^\prime
l^\pprime}$ for $l=0$ can be easily obtained from the formulae in 
Appendix A.
\be
\left.\sqrt{l(l+1)}\alpha^{m m^\prime m^\pprime}_{l l^\prime l^\pprime}
\right|_{l=0} ={(-1)^{m^\prime}\over \sqrt{4\pi}}
\delta_{l^\prime,l^\pprime}\delta_{m^\prime-m^\pprime},
\ee
and using the relation (\ref{realcoe}) between coefficients of opposite $m$, 
eq.~(\ref{radfor2}) for $l=0$ simplifies to
\be
\sum_{lm} |A_{lm}|^2=4\pi H_0(1+H_0),
\label{normA}
\ee
whereas, for $l\ge 1$, it reads
\be
{H_0\over \sqrt{l(l+1)}} R_{lm}=
\sum_{l^\prime m^\prime}\;\sum_{l^\pprime m^\pprime}
\left[{l^\prime(l^\prime+1)\over l^\pprime(l^\pprime+1)}\right]^{1/2}
\alpha^{m m^\prime m^\pprime}_{l l^\prime l^\pprime}
A_{l^\prime m^\prime}A_{l^\pprime m^\pprime}.
\label{radfor3}
\ee

We now expand the equation for the tangential force, eq.~(\ref{tanfor})
along the two orthogonal sets of vectors ${\bf B}_{lm}$ and ${\bf C}_{lm}$.
With the expansions given above, eq.~(\ref{tanfor}) becomes
\[
\sum_{l^\prime m^\prime}{l^\prime (l^\prime +1)-2(1+H_0)\over 2\sqrt{l^\prime(l^\prime +1)}}
R_{l^\prime m^\prime }{\bf B}_{l^\prime m^\prime}
+\sum_{l^\prime m^\prime}\sum_{l^\pprime m^\pprime}
\sqrt{l^\pprime(l^\pprime+1)}
R_{l^\prime m^\prime}V_{l^\pprime m^\pprime}
Y_{l^\prime m^\prime}{\bf B}_{l^\pprime m^\pprime}  
\]
\be
+\sum_{l^\prime m^\prime}\sum_{l^\pprime m^\pprime}
\sqrt{l^\pprime(l^\pprime+1)}
A_{l^\prime m^\prime}A_{l^\pprime m^\pprime}
Y_{l^\prime m^\prime}{\bf B}_{l^\pprime m^\pprime}=0.
\label{tanfor1}
\ee
We take the scalar product of this equation with ${\bf B}^\ast_{lm}$ and
integrate over solid angle $d\Omega$, using the relations eq.~(\ref{Vlm})
and (\ref{Flm}) and defining the coupling coefficients $\beta^{m m^\prime
m^\pprime}_{l l^\prime l^\pprime}$ (see Appendix~A),
\be
\beta^{m m^\prime m^\pprime}_{ll^\prime l^\pprime}\equiv {1\over \sqrt{l^\prime(l^\prime+1)}}\oint
Y_{l^\prime m^\prime }{\bf B}_{l^\pprime m^\pprime}\cdot {\bf B}^\ast_{lm}\;d\Omega.
\label{beta}
\ee
The result is 
\[
{l(l+1)-2(1+H_0)\over 2\sqrt{l(l+1)}} R_{lm}
-\sum_{l^\prime m^\prime}\sum_{l^\pprime m^\pprime}
\left[{l^\prime(l^\prime+1)\over l^\pprime(l^\pprime+1)}\right]^{1/2}
\beta^{m m^\prime m^\pprime}_{ll^\prime l^\pprime}
R_{l^\prime m^\prime}R_{l^\pprime m^\pprime}
\]
\be
+\sum_{l^\prime m^\prime}\sum_{l^\pprime m^\pprime}
[l^\prime(l^\prime+1)l^\pprime(l^\pprime+1)]^{1/2}
\beta^{m m^\prime m^\pprime}_{ll^\prime l^\pprime}
A_{l^\prime m^\prime}A_{l^\pprime m^\pprime}=0.
\label{expans_b}
\ee

We have noticed in Sect.~4 that the tangential component of the equation
of force balance has no component in the direction of electric-current
lines.  Since we see from eq.~(\ref{jcurr}) that ${\bf j}$ is expressed
as a series containing the harmonics ${\bf C}_{lm}$, we expect therefore
each coefficient of the expansion of eq.~(\ref{tanfor1}) in terms of
${\bf C}_{lm}$ vectors to be zero. This is verified in Appendix~B.

Summarizing, the equations of the problem are:
\be
{H_0\over \sqrt{l(l+1)}} R_{lm}=
\sum_{l^\prime m^\prime}\;\sum_{l^\pprime m^\pprime}
\left[{l^\prime(l^\prime+1)\over l^\pprime(l^\pprime+1)}\right]^{1/2}
\alpha^{m m^\prime m^\pprime}_{l l^\prime l^\pprime}
A_{l^\prime m^\prime}A_{l^\pprime m^\pprime},
\ee
from the condition of force balance in the radial direction;
\[
{l(l+1)-2(1+H_0)\over 2\sqrt{l(l+1)}} R_{lm}
=
-\sum_{l^\prime m^\prime}\sum_{l^\pprime m^\pprime}
[l^\prime(l^\prime+1)l^\pprime(l^\pprime+1)]^{1/2}
\beta^{m m^\prime m^\pprime}_{ll^\prime l^\pprime}
A_{l^\prime m^\prime}A_{l^\pprime m^\pprime}
\]
\be
+\sum_{l^\prime m^\prime}\sum_{l^\pprime m^\pprime}
\left[{l^\prime(l^\prime+1)\over l^\pprime(l^\pprime+1)}\right]^{1/2}
\beta^{m m^\prime m^\pprime}_{ll^\prime l^\pprime}
R_{l^\prime m^\prime}R_{l^\pprime m^\pprime},
\ee
from the condition of force balance of force in the tangential direction;
and  
\be
\sum_{lm} |A_{lm}|^2=4\pi H_0(1+H_0),
\ee
defining the amount of support provided by the magnetic field (monopole
component of the equation of force balance in the radial direction).
In this way, the representation of the physical variables has been
transferred from function space, in terms of $\theta$ and $\varphi$, to
the infinite-dimensional Hilbert space, each vector component now being
the coefficient of the corresponding harmonic. The process is analogous
to transforming from the Schr\"odinger to the Heisenberg description in
quantum mechanics (it is incomplete, in that the radial dependence is
still represented in function space).

The procedure for solving the equations of MS in spectral form is to
select a finite set of coefficients by truncating the series expansion
to some $l=l_{\rm max}$, setting to zero all remaining coefficients.
As a test of the method, in the next two subsections we solve the MS
equations with $m=0$ for various values of $l_{\rm max}$, and compare
the results with the axisymmetric solutions obtained by MN00 and LS96
for the hydrostatic and magnetostatic case, respectively.

\section{Hydrostatic equilibria}

Setting $H_0=0$ and $A_{lm}=0$ for any $(l,m)$, we obtain the set of
equations governing hydrostatic equilibria, 
\be
{l(l+1)-2\over 2\sqrt{l(l+1)}} R_{lm}
=\sum_{l^\prime}\sum_{l^\pprime} 
\left[{l^\prime(l^\prime+1)\over l^\pprime(l^\pprime+1)}\right]^{1/2}
\beta_{ll^\prime l^\pprime}^{mm^\prime m^\pprime}
R_{l^\prime m^\prime} R_{l^\pprime m^\pprime}.
\label{nonmag}
\ee
Remarkably, for $l=1$ both the LHS and RHS of eq.~(\ref{nonmag}) are
zero. To see this, observe that for $l=1$ the triangular relation imply
that all coefficients $\beta_{ll^\prime l^\pprime}^{mm^\prime m^\pprime}$
vanish unless $l^\prime=l^\pprime$, or $l^\prime=l^\pprime\pm 1$. Thus,
the RHS contains terms like
\be
\beta_{1l^\prime l^\prime}^{m m^\prime m^\prime}R_{l^\prime m^\prime}^2,
\label{t1}
\ee
or like
\be
\left\{\left[{l^\prime (l^\prime +1)\over 
(l^\prime +1)(l^\prime +2)}\right]^{1/2}
\beta_{1l^\prime l^\prime +1}^{m m^\prime m^\pprime}+
\left[{(l^\prime +1)(l^\prime +2)\over l^\prime (^\prime l+1)}\right]^{1/2}
\beta_{1l^\prime +1l^\prime }^{m m^\pprime m^\prime}\right\}
R_{l^\prime m^\prime}  R_{l^\prime +1 m^\pprime}.
\label{t2}
\ee
Terms like (\ref{t1}) are zero because $\alpha$ and $\beta$
coefficients are zero when $l+l^\prime+l^\pprime$ is odd (see Appendix~A);
terms like (\ref{t2}) are zero because 
\be
\beta^{m m^\prime m^\pprime}_{1l^\prime l^\prime+1}=
-{l^\prime+2\over l^\prime}
\beta^{m m^\pprime m^\prime}_{1l^\prime+1 l^\prime},
\ee
as shown in Appendix~A. Thus, all terms allowed by the triangular
conditions in eq.~(\ref{nonmag}) with $l=1$, are identically zero. This
implies that the coefficient $R_{1m}$ (or $V_{1m}$) is undefined,
an intrinsic degree of freedom of the problem that we refer to as the
``dipole gauge''.  The expansion of the density function $R$ contains
dipole terms proportional to $Y_{1m}(\theta,\varphi)$ with $m=0,\pm 1$,
\be
R(\theta,\varphi)=1+\sum_{m=-1,0,1}R_{1m}Y_{1m}(\theta,\varphi)+
\ldots=1+(c_1\cos\theta+c_2\cos\theta\cos\varphi+c_3\cos\theta\sin\varphi)
+\ldots,
\label{Rseries}
\ee
where $c_1$, $c_2$ and $c_3$ are real coefficients. The last three
terms represent an eccentric distortion of the basic spherically
symmetric equilibrium ($R=1$, the singular isothermal sphere) along
three perpendicular axes, confirming the result of MN00 that the singular
isothermal sphere is neutrally stable with respect to dipole-like density
perturbations. Computing additional terms in the series (\ref{Rseries}),
the series converges to the function $R(\theta,\varphi)$ for each value of
$R_{1m}$.  Without loss of generality we set $m=0$, equivalent to assuming
that the symmetry axis of the configuration lies in the $z$ direction,
and for better clarity we omit the index 0 in the expansion terms and
coupling coefficients.
For $l_{\rm max}=2$, we have to solve for $R_{2}$ as function of $R_{1}$
the quadratic equation
\be
2 R_{2}=\sqrt{6}\beta_{111} R_{1}^2
+(\sqrt{2}\beta_{212}+3\sqrt{2}\beta_{221}) R_{1} R_{2} +\sqrt{6}R_{2}^2,
\ee
with solution
\be
R_{2}={14\pi\over 5}
\left[1-\sqrt{1-\left({3\over 28\pi}\right)R_1^2}\right]\simeq
{3\over 20}R_{1}^2 \qquad \mbox{if $R_{1}\ll 1$.} 
\ee
The fast decrease of the coefficients with increasing $l$, at least for
small $R_{1}$, suggests a rapid convergence of the series expansion.
For $l_{\rm max}\ge 2$, the system of truncated equations can be easily
solved by iteration. The procedure usually converges to a solution with
a relative accuracy of less than $10^{-5}$ in 4 iterations.  The series
solution obtained in this way corresponds to the ellipsoidal solutions
parametrized by the ellipticity $e$ (see Sect.~3.3), found by MN00.
We show in Fig.~\ref{h0m0} the solution obtained assuming $R_1=2$
for $l_{\rm max}=2,3$ and 4. As anticipated, the terms of the series
expansion decrease exponentially with increasing $l$.  For comparison we
also show the corresponding analytical solution of MN00 (eq.~[\ref{mn}],
with $e=0.466$)

In principle, the sequence of equilibria defined by the parameter
$R_1$, with $0<R_1<\infty$, can bifurcate into non-axisymmetric 
configurations of equilibrium. A linearization of eq.~(\ref{nonmag}) 
gives the condition
\be
{l(l+1)-2\over 2\sqrt{l(l+1)}} 
- \left\{\left[{2\over l(l+1)}\right]^{1/2}\beta_{l1l}^{m0m}+
\left[{l(l+1)\over 2}\right]^{1/2}\beta_{ll1}^{mm0}\right\}R_{1}=0,
\ee
for the occurrence of bifurcations with angular dependence defined by
harmonics with general $l,m$ along the $l=1,m=0$ sequence.  However,
$\beta_{l1l}^{m0m}=\beta_{ll1}^{mm0}=0$ since $2l+1$ is odd (see Appendix
A), and therefore no azimutally asymmetric bifurcation occurs along the
sequence of hydrostatic equilibria generated by the $R_{1}$ term.

\section{Magnetostatic equilibria with azimuthal and equatorial symmetry}

In this section we look for highly symmetrical solutions, possessing
both azimuthal and equatorial symmetry. We thus set $m=0$, and we 
impose the existence of a plane of symmetry at $\theta=\pi/2$
(equatorial plane). The density $R$, the gravitational potential $V$, the
components $B_\theta$ and $B_\varphi$ of the magnetic field are symmetric
with respect to the equatorial plane, whereas the radial component of
the magnetic field, $B_r$, is anti-symmetric.  This implies that the
expansion of $V$ and $R$ contains multipole terms with $l$ even, whereas
the expansion of $A$ (and $F$) contains multipole terms with $l$ odd.
Since $m=0$, all expansion coefficients are real. As in the previous 
section, we omit the index $m$ in the expansion terms and coupling
coefficients.

We truncate the infinite system of nonlinear algebraic equations to some
index $l_{\rm max}$ and we set all coefficient of the expansion equal to
zero for $l>l_{\rm max}$.  At the lowest level of approximation, $l_{\rm
max}=2$, we have to solve two equations for the two coefficients $A_1$
and $R_2$,
\be
{H_0\over\sqrt{6}}R_2=\alpha_{211}A_1^2,
\ee
\be
A_1^2=4\pi H_0(1+H_0),
\ee
where $\alpha_{211}=-1/\sqrt{120\pi}$. The solution is 
\be
R_2=-2\sqrt{\pi\over 5}(1+H_0).
\qquad 
A_1=\sqrt{4\pi H_0(1+H_0)},
\label{ansol}
\ee
The coefficients $R_2$ and $A_1$ determined in this way constitute 
the lowest-order terms of a series expansion for the
the functions $R(\theta)$ and $\phi(\theta)$,
\be
R(\theta)\approx {3\over 2}(1+H_0)\sin^2\theta, 
\qquad
\phi(\theta)\approx {1\over 2}\sqrt{3H_0(1+H_0)}\sin^2\theta, 
\ee
and the spherical mass-to-flux ratio, from eq.~(\ref{lambdar}),
\be
\lambda_r\approx {4(1+H_0)\over \phi_1 A_1}=2\sqrt{1+H_0\over 3H_0}.
\label{lambdar_an}
\ee
This series solution converges to the solution determined by LS96
(singular isothermal toroids).  As in the case of the hydrostatic
equilibria discussed in the previous section, the convergence of the
series solution is rapid: the approximated expression of $\lambda_r$,
obtained with only the first term in the expansion of $R(\theta)$ and
$\phi(\theta)$, is in good agreement with the numerical values calculated
by LS96 as function of $H_0$, the largest discrepancy being $\sim 15\%$
for $H_0 \gg 1$.

To obtain a more accurate series solution, for $l_{\rm max}>2$ we solve
numerically the equations of MS in spectral form with Newton's method,
assuming as a first guess the analytical solution eq.~(\ref{ansol}). The
procedure converges to a solution with a relative accuracy of $10^{-8}$
in 4 iterations.  In Table~1 we list the values of the coefficients $A_l$
and $R_l$, and the spherical mass-to-flux ratio for $H_0=0.5$ and $l_{\rm
max}=2,3,4,5$ and 6.  The comparison with the exact numerical solution
of LS96 is shown in Fig.~\ref{h05m0_a} and \ref{h05m0_b}.

\section{Three-dimensional magnetostatic equilibria}

Useful information about the existence of solutions of the MS equations
with lower degrees of symmetry than those considered in Sect.~5 and 6, can
be obtained by linearization of the spectral equations.  The occurrence
of neutrally stable configurations along the sequence of equilibria
controlled by the parameter $H_0$ is indicated by the vanishing of
the determinant of the linearized system for some value of the control
parameter (in our case the parameter $H_0$ measuring the relative amount
of support provided by the magnetic field).  The zeroes of the determinant
signal the presence of neutrally stable equilibria or bifurcations points
in the sense of Poincar\'e (see e.g. Galli et al.~2001).

In this way, we show in Sect.~7.1 and 7.2 that the degree of freedom
represented by the ``dipole gauge'' affecting hydrostatic equilibria
(Sect.~5), is also present in the case of MS equilibria independently
on the degree of support provided by the magnetic field.  The magnetic
field however removes the degeneracy of the purely hydrostatic case,
where the density distortion with $l=1$ and $m=0,\pm 1$ gives origin to
three orthogonal orientations of the {\it same} configuration.  Magnetized
equilibria instead are split by the ``dipole gauge'' into two families:
one possessing an equatorial plane of symmetry but azimutally asymmetric
(with $m=\pm 1$) and elongated in two orthogonal directions; and one
without a plane of symmetry but azimuthally symmetric (with $m=0$).
In addition, the dipolar distortion of the density is coupled to a
quadrupolar distortion of the vector potential, with the same $m$.

Finally, in Sect.~7.3 we address the question of the existence of
neighboring equilibria to the basic axisymmetric solutions determined in
Sect.~6, limiting the analysis to perturbations in the density described
by sectorial harmonics (with $l=m$).  We find that the only allowed
perturbation of this kind has $l=m=1$, as anticipated in Sect.~7.2.

\subsection{Case $l=1, m=0$: equatorially asymmetric density distortions}

First we consider axisymmetric distortions that introduce an asymmetry
of the configuration with respect to the equatorial plane. These are
represented by the small coefficients $R_{10}$ for the density and
$A_{20}$ for the vector potential, and are governed by the linearized
equations
\be
\sqrt{3}H_0R_{10}-\sqrt{2}(\alpha_{112}^{000}+3\alpha_{121}^{000})
\tilde A_{10}A_{20}=0,
\ee
and
\be
\sqrt{3}H_0R_{10}+\sqrt{2}(\beta_{112}^{000}+3\beta_{121}^{000})
\tilde R_{20}R_{10}
-6\sqrt{2}(\beta_{112}^{000}+\beta_{121}^{000})\tilde A_{10}A_{20}=0,
\ee
where a tilde ($\;\tilde{}\;$) indicates the coefficients of the
axisymmetric solution.

A bifurcation can occur if the determinant of the system is zero.
Substituting the values of the coefficients, we see that the derminant
vanishes for any value of $H_0$. For these linearized equilibria,
the relation between $R_{10}$ and $A_{20}$ is
\be
H_0\sqrt{5\pi\over 2}R_{10}=\tilde A_{10}A_{20}.
\ee
An example of these azimuthally symmetric equilibria lacking equatorial
symmetry is shown in Fig.~\ref{l1} (left panel).

\subsection{Case $l=1,m=\pm 1$: azimuthally asymmetric density distortions}

Next, we consider density perturbations azimuthally asymmetric 
but conserving the symmetry of the original equilibrium state with 
respect to the equatorial plane. These are characterized by expansion
coefficients $R_{11}$ and  $A_{21}$ for the density and the vector
potential, respectively, and are governed by the linearized equations 
\be
\sqrt{3}H_0R_{11}-\sqrt{2}(\alpha_{112}^{101}+3\alpha_{121}^{110})\tilde A_{10}A_{21}=0,
\ee
and
\be
\sqrt{3}H_0R_{11}+\sqrt{2}(\beta_{112}^{110}+3\beta_{121}^{101})\tilde R_{20}R_{11}
-6\sqrt{2}(\beta_{112}^{101}+\beta_{121}^{110})\tilde A_{10}A_{21}=0.
\ee
As before, the coefficients $\tilde R_{20}$ and $\tilde A_{10}$ are
those of the axisymmetric solution. 

A bifurcation can occur if the determinant of the system is zero for
some value of $H_0$.  Substituting the values of the coefficients,
we see that the determinant vanishes for any value of $H_0$. For these
linearized equilibria, the relation between $R_{11}$ and $A_{21}$ is
\be 
H_0\sqrt{5\pi\over 3}R_{11}=\tilde A_{10}A_{21}.
\ee
An example of these azimuthally asymmetric equilibria with equatorial
symmetry is shown in Fig.~\ref{l1} (right panel).

\subsection{Sectorial density distortions: linear analysis}

To assess the validity of Parker's theorem for axisymmetric
self-gravitating equilibria, we check whether the class of axisymmetric
solutions obtained in Sect.~6 (converging to the singular isothermal
toroids of LS96) allows neighbouring 3-D equilibria. To this goal, we
perform a linearization of the set of nonlinear algebraic equations near
the axisymmetric solution and we consider for simplicity distortions of
the density function with arbitrary $l$, assuming for simplicity $m=l$
(sectorial distortions).  As before, the expansion coefficients
corresponding to the axisymmetric solution are indicated by a tilde
($\;\tilde {}\;$).

The linearized equations for small distortions characterized by 
expansion coefficients $R_{ll}$ and $A_{l+1l}$, with arbitrary $l$, are 
\be
\frac{H_0}{\sqrt{l(l+1)}}R_{ll}=\left(\sqrt{\frac{2}{(l+1)(l+2)}}
\alpha_{l1l+1}^{l0l}+\sqrt{\frac{(l+1)(l+2)}{2}}
\alpha_{ll+11}^{ll0}\right)\tilde A_{10}A_{l+1l},
\ee
\begin{eqnarray}
\frac{l(l+1)-2(1+H_0)}{2\sqrt{l(l+1)}}R_{ll} & = &
\left(\sqrt{\frac{6}{l(l+1)}}
\beta_{l2l}^{l0l}+\sqrt{\frac{l(l+1)}{6}}\beta_{ll2}^{ll0}\right)
\tilde R_{20}R_{ll} \nonumber \\
& & 
-\sqrt{2(l+1)(l+2)}(\beta_{l1l+1}^{l0l}+\beta_{ll+11}^{ll0})
\tilde A_{10}A_{l+1l}.
\end{eqnarray}

The determinant $\Delta_{ll}$ of this systems of equations is 
\begin{eqnarray}
\Delta_{ll}/\tilde A_{10} & = 
& H_0\sqrt{\frac{2(l+2)}{l}}(\beta_{l1l+1}^{l0l}+\beta_{ll+11}^{ll0})
+\left(\sqrt{\frac{2}{(l+1)(l+2)}}\alpha_{l1l+1}^{l0l}
+\sqrt{\frac{(l+1)(l+2)}{2}}\alpha_{ll+11}^{ll0}\right) \nonumber \\
& & 
\left[\frac{l(l+1)-2(1+H_0)}{2\sqrt{l(l+1)}}-\left(\sqrt{\frac{6}{l(l+1)}}
\beta_{l2l}^{l0l}+\sqrt{\frac{l(l+1)}{6}}\beta_{ll2}^{ll0}\right)
\tilde R_{20}\right],
\end{eqnarray}
and vanishes for $l=1, m=1$, as anticipated in Sect.5 and 7.2.
For $l\neq 1$, we have evaluated the determinant numerically as
function of $H_0$ for $l=2,3,4,5$ and 6, and $0<H_0<2$. The results
are shown in Fig.~\ref{det}. At least in this part of parameter space,
the determinant is always positive, its value monotonically increasing
with $H_0$ and $l$, and shows no sign of having a zero for particular
values of $H_0$.  We can then safely conclude that the axisymmetric MS
equilibria determined by LS96 and discussed in Sect.~6, have no other
neighboring equilibria than those allowed for all values of $H_0$ 
by the ``dipole gauge'' discussed in Sect.~6, 7.1 and 7.2.

In this sense, these non-symmetric equilibria represent the only
exceptions (within the assumption of the present study) to a generalized
version of Parker's theorem, originally formulated for systems with
translational symmetry, extended to self-gravitating equilibria with axial
symmetry. In the next section we consider the equilibria originated by
the ``dipole gauge'' from a physical point of view, and we conclude that
{\it all} these solutions are probably not force-free at the origin,
a singular point for a scale-free configuration, and therefore they
cannot represent realistic models of equilibrium of isolated cosmic bodies.

\section{Summary}

The results obtained in this paper show that previously known axisymmetric
solutions of the MS equations for an isothermal self-gravitating
gas, under the hypothesis of scale invariance and global neutrality,
allow only neighbouring equilibria characterized by a $l=1, m=0,\pm 1$
angular dependence of the density distortion for {\it any} value of the
degree of magnetic support (including zero and infinite).  For $m=0$,
the original axisymmetric equilibrium is distorted by a ``bending''
of the isodensity contours with respect to the equatorial plane,
preserving azimuthal symmetry; for $m=\pm 1$, the equilibrium is
distorted by a ``stretching'' of the isodensity contours along one of
two orthogonal directions in the equatorial plane, preserving up-down
reflection symmetry.  In the absence of magnetic fields (for $H_0=0$)
these two classes of distorted equilibria reduce to the ellipsoidal
equilibria found by MN00. In the limit of vanishing thermal support
($H_0\rightarrow\infty$) the configuration of equilibrium reduces to a
thin disk supported only by magnetic tension against its self-gravity,
and the $m=1$ density distortion corresponds to the elliptical disklike
equilibria found by Galli et al.~(2001). What is the significance of
the neutral stability of these configurations to density distortions
characterized by a dipolar angular dependence?

According to MN00, the ellipsoidal hydrostatic equilibria are not
force-free at the origin, where the gravity of all the matter of the
configuration produces a non-vanishing force trying to restore symmetry
with respect to the equatorial plane. According to Cai \& Shu~(2004),
the same conclusion holds for the azimutally asymmetric solutions found
by Galli et al.~(2001) for MS equilibria in the thin-disk limit. These
two classes of equilibria stem from the singular isothermal sphere and
the singular isothermal disk, respectively, that in turn are the limiting
cases (for $H_0=0$ and $H_0=\infty$) of the family of singular isothermal
toroids of LS96.  It is therefore tempting to conclude, that all singular
isothermal equilibria, characterized by a distortion of the isodensity
(or equipotential) surfaces proportional to the $l=1$ harmonics, are not
force-free at the origin for any value of $H_0$. If this is the case,
their relevance to represent realistic equilibria is doubtful.

The same conclusion probably hold for configurations rotating with
spatially uniform velocity $u_\varphi$ (the only rotation law compatible
with isothermality and spatial self-similarity). Galli et al.~(2001)
found that rotating singular isothermal disks are neutrally stable to
$m=1$ perturbations for any value of the rotation rate. In a similar vein,
MN00 found that the the azimuthally and equatorially symmetric rotating
models of Toomre~(1982) and Hayashi et al.~(1982) can be ``continued'',
for any value of the rotation velocity, into a sequence of axisymmetric
equilibria lacking equatorial symmetry (in our language, originated by
a $l=1,m=0$ density distortion). At the light of the results described
in Sect~7.1 and 7.2, it is natural to expect that the sequence of
Toomre-Hayashi models also possess non-axisymmetric counterparts with
a dominant $l=1,m=1$ asymmetry, although solutions of this kind are not
known.  Likely, also these hypothetical rotating asymmetric equilibria
are not force-free at the origin. The $l=1$, $m=0,\pm 1$ distortion,
the allowed distortion of axisymmetric equilibria found in the present
study, may then represent a {\it gauge freedom} that creeps somehow into
general self-similar isothermal equilibria, irrespectively of the presence
of magnetic fields, or rotation, or else.  The appearance of this gauge
freedom in self-similar isothermal systems, and its physical significance,
deserves further scrutiny. As a counterexample, it should be easy to show
that, assuming a non-isothermal (e.g. polytropic) equation of state,
the analogues of the Li-Shu magnetized equilibria studied by Galli et
al.~(1999), (or the polytropic analogues of the Toomre-Hayashi models)
are not affected by this gauge freedom.

For arbitrary $l=m\neq 1$, a perturbation analysis shows that
the symmetric magnetized solutions found by LS96 do {\it not} have
neighbouring non-axisymmetric equilibria. This results is analogous to
Parker's theorem for systems with translational invariance, and suggests
that the validity of Parker's theorem can be extended to self-gravitating
axisymmetric equilibria. Combining this result with the findings of
Galli et al.~(2001), one is led to the conclusion that the presence of
fluid motions (specifically, super-magnetosonic rotation) is a crucial
ingredient for the occurrence of symmetry break-ups in MHD equilibria.

The results of this paper imply that, {\it under the assumed
conditions}, very few (and possibly not physically meaningful)
non-axisymmetric solutions of the steady MHD equations do exist. Of the
assumed conditions, probably the most severe are the assumptions that
({\it i}\/) the magnetic field is analytic everywhere, and that ({\it
ii}\/) the new, non-symmetric solutions are accessible from the basic
states by regular perturbations (i.e. small-parameter expansions). As
for the former assumption, Moffatt~(1985) has shown that magnetostatic
equilibria of non-trivial topology in a perfectly conducting fluid
may generally contain tangential discontinuities (current sheets),
i.e. they cannot be described in terms of analytic functions; in
contrast, equilibrium fields that are analytic functions of space are
subject to severe structural constraints, as shown by Arnol'd~(1965,
1966). As for the latter assumption, our results show that equilibria
of non-trivial topology, if they exist, cannot in general be reached
starting from axisymmetric states by regular perturbation. This is consistent
with the conclusions of Rosner \& Knobloch~(1982), that the response of stationary 
solutions of nonlinear equations to finite-amplitude, symmetry-breaking
perturbations may not in general be obtained in terms of small-parameter
expansions of the variables. These limitations should be kept in mind in 
interpreting the results of this paper. 

\subsection{Implications for molecular cloud cores}

Recent statistical studies (Jones, Basu, \& Dubinski 2001; Jones \&
Basu 2002; Goodwin, Ward-Thompson, \& Whitworth 2003) based on
available catalogues of molecular cloud cores and Bok globules (typical
size $L=0.1$~pc, sound speed $c_{\rm s}=0.2$~km~s$^{-1}$, average
density $n({\rm H}_2)=10^4$~cm$^{-3}$, and typical magnetic field
strength $B=10$~$\mu$G), do not support the possibility that cores are
axisymmetric configurations. A good fit to the observed axial
distribution is generally found assuming instead that cores are
triaxial ellipsoids.  The best-fit axial ratios $a:b:c$ determined
statistically for molecular cloud cores ($1:0.9\pm 0.1:0.5\pm 0.1$
according to Jones \& Basu~2002; $1:0.8\pm 0.2:0.4\pm 0.2$ according to
Goodwin, Ward-Thompson, \& Whitworth~2003), suggest that cores are
preferentially flattened in one direction and nearly oblate ($a\approx
b>c$), and imply that they may not be particularly far from conditions
of equilibrium.

Taking these observational results at face value, one is led to
consider the fate of a cosmic cloud with its frozen-in magnetic field
formed by whatever process in a configuration lacking a high degree of
symmetry and presumably not in an exact equilibrium state.  As
discussed in Sect.~2.2, Parker (1979) argued that realistic magnetic
fields with no well defined symmetries must evolve in a genuinely
time-dependent way, until all non-symmetric components of the field are
destroyed by dissipation and reconnection and the topology becomes
symmetric.  This process can hardly be of any relevance for the
interstellar gas, where ohmic dissipation times are larger than the age
of the Universe.  Ambipolar diffusion on the other hand can only
redistribute the mass inside flux tubes and drive a stable equilibrium
to the threshold of dynamical instability, but cannot dissipate the
magnetic energy stored in the field.

To the extent that interstellar clouds can be represented as isolated
MS equilibria, these systems must instead undergo Alfv\`en oscillations
(weakly damped by ambipolar diffusion) around the equilibrium state
with period (Woltjer~1962)
\be
\tau\approx {L\over (c_{\rm s}^2 +v_{\rm A}^2)^{1/2}},
\ee
where $L$ is the size of the system, $c_{\rm s}$ is the sound speed and
$v_{\rm A}$ is the Alfv\`en speed.  This kind of behaviour is
evident in the numerical and analytical calculations of
Hennebelle~(2003) relative to the homologous evolution of prolate and oblate
magnetized isothermal spheroids.  In the non-selfgravitating case, the
stability of magnetostatic equilibria of arbitrary complex topology was
studied by Moffatt~(1986) through construction of the second variations
of the magnetic and kinetic energies with respect to a virtual
displacement about the equilibrium configuration. Moffatt's~(1986)
results show that a general class of space-periodic magnetostatic
equilibria is stable to disturbances of arbitrary lengthscale. If
perturbed in some way, the fluid responds executing oscillations about
this equilibrium, that are eventually damped if due account is taken of
viscosity.
If we adopt the deviation from axial symmetry of the shapes of cloud
cores as rough measure of their nonequilibrium, a simple harmonic
oscillator analogy provides an estimate of the average velocity
$\langle v^2\rangle^{1/2}$ of pulsation,
\be 
\frac{\langle v^2\rangle}{(c_{\rm s}^2+v_{\rm A}^2)} 
\approx \frac{\langle a-b\rangle^2}{a^2}, 
\ee 
that, for $a:b=1:0.8$ gives $\langle v^2\rangle^{1/2}\approx
0.2\;({c_{\rm s}^2+v_A^2})^{1/2}$. For typical conditions of molecular
cloud cores, this implies coherent pulsations with period $\tau \approx
4\times 10^5$~yr, average velocity $\langle v^2\rangle^{1/2}\approx
0.05$~km~s$^{-1}$, and maximum velocity $\sim 0.1$~km~s$^{-1}$. These
pulsational motions may have already been detected in the isolated
globule B68 (Lada et al.~2003).

\bigskip

\acknowledgements
It is a pleasure and a privilege to thank Frank Shu, who proposed the
problem, suggested the solution technique, and anticipated the results,
long before the calculations presented in this paper were completed.
An anonymous referee is also aknowledged for a very insightful report
that helped to improve the original manuscript.

\clearpage

\appendix 

\section{The coupling coefficients}

The coupling integrals of eq.~(\ref{alpha}) and (\ref{beta}) can be
expressed in terms of integrals of three spherical harmonics using
standard recurrence formulae to eliminate the derivatives. The
resulting integrals can be evaluated with the Gaunt formula,
\be
\oint Y^\ast_{lm} Y_{l^\prime m^\prime} Y_{l^\pprime m^\pprime}
\; d\Omega=[l(l+1)l^\prime(l^\prime+1)l^\pprime(l^\pprime+1)]^{1/2}
N_{ll^\prime l^\pprime}G^{m m^\prime m^\pprime}_{ll^\prime l^\pprime},
\ee
where
\be
N_{ll^\prime l^\pprime}\equiv
{1\over 2}\left[{(2l+1)(2l^\prime+1)(2l^\pprime+1)
\over 4\pi l(l+1)l^\prime(l^\prime+1)l^\pprime(l^\pprime+1)}
\right]^{1/2},
\ee
and 
\be
G^{m m^\prime m^\pprime}_{l l^\prime l^\pprime}\equiv
(-1)^m\left(
\begin{array}{ccc}
l & l^\prime & l^\pprime \\
0 & 0        & 0
\end{array}
\right)
\left(
\begin{array}{ccc}
l  & l^\prime & l^\pprime \\
-m & m^\prime & m^\pprime
\end{array}
\right).
\label{Gaunt}
\ee
The $3j$ symbols are algebraically defined by e.g. the Racah formula
(see e.g. Landau \& Lifshitz~1997, Varshalovich, Moskalev 
\& Khersonskii~1998).  
The evaluation of the coupling
coefficients is best achieved by group-theoretical methods
and the use of Wigner $6j$ and $3j$ symbols. Here we quote only the
results, details on the procedure can be found in Jones~(1985).
In terms of the quantities $N_{ll^\prime l^\pprime}$ and 
$G^{m m^\prime m^\pprime}_{ll^\prime l^\pprime}$, we have 
\be
\alpha^{m m^\prime m^\pprime}_{ll^\prime l^\pprime}=\Lambda_l N_{ll^\prime l^\pprime}
G^{m m^\prime m^\pprime}_{ll^\prime l^\pprime},\qquad 
\beta^{m m^\prime m^\pprime}_{ll^\prime l^\pprime}=\Lambda_{l^\prime} N_{ll^\prime l^\pprime}
G^{m m^\prime m^\pprime}_{ll^\prime l^\pprime},
\ee
where 
\be
\Lambda_l=l^\pprime(l^\pprime+1)+l^\prime(l^\prime+1)-l(l+1),\qquad
\Lambda_{l^\prime}=l^\pprime(l^\pprime+1)+l(l+1)-l^\prime(l^\prime+1).
\ee
The coupling coefficients $\alpha^{m m^\prime m^\pprime}_{ll^\prime
l^\pprime}$ and $\beta^{m m^\prime m^\pprime}_{ll^\prime l^\pprime}$
are real, and are equal to zero when the following (triangular) 
conditions on the Wigner $3j$ symbols are not all satisfied:
\be
l\le l^\prime+l^\pprime, \qquad
l^\prime \le l+l^\pprime, \qquad
l^\pprime \le l+l^\prime, \qquad l+l^\prime+l^\pprime\;\mbox{even}, 
\label{triang_l}
\ee
and
\be
m=m^\prime+m^\pprime.
\label{triang_m}
\ee
We recall that a $3j$ symbol is invariant under even
permutation of columns and is multiplied by the phase factor
$(-1)^{(l+l^\prime+l^\pprime)}$ for odd permutations. Thus $G^{m m^\prime
m^\pprime}_{ll^\prime l^\pprime}$ being the product of two $3j$ symbols,
is invariant for any permutation.  Using this property, one obtains,
for example,
\be
\beta^{m m^\prime m^\pprime}_{1ll+1}=
-{l+2\over l}\beta^{m m^\pprime m^\prime}_{1l+1l}.
\ee

The coefficients $\alpha^{m m^\prime m^\pprime}_{ll^\prime l^\pprime}$ 
and $\beta^{m m^\prime m^\pprime}_{ll^\prime l^\pprime}$ are real, whereas 
$\gamma^{m m^\prime m^\pprime}_{l l^\prime l^\pprime}$ is imaginary:
\be
\gamma^{m m^\prime m^\pprime}_{l l^\prime l^\pprime}=-i\Gamma_l N_{ll^\prime l^\pprime} 
H^{m m^\prime m^\pprime}_{ll^\prime l^\pprime},
\ee
where 
\be
H^{m m^\prime m^\pprime}_{l l^\prime l^\pprime}=
(-1)^m \left(
\begin{array}{ccc}
l & l^\pprime & l^\prime-1\\
0 & 0         & 0
\end{array}
\right)
\left(
\begin{array}{ccc}
l & l^\prime & l^\pprime\\
-m & m^\prime & m^\pprime 
\end{array}
\right),
\ee
and
\be
\Gamma_l=[(l^\prime+l^\pprime-l)(l^\prime+l-l^\pprime)
(-l^\prime+l+l^\pprime+1)(l^\pprime+l^\prime+l+1)]^{1/2}.
\ee
Notice that all the dependence of the coupling coefficients on the
azimuthal number $m$ is contained in the Wigner $3j$ symbols, whereas
the remaining factors depend on $l$ only (the Wigner-Eckart theorem).

\section{Proof that one component of the equation of force balance is
trivially satisfied}

To show this, we take then the scalar product of eq.~(\ref{tanfor1})
with ${\bf C}^\ast_{lm}$ and integrate over solid angle, obtaining the
condition
\be
\sum_{l^\prime m^\prime}\sum_{l^\pprime m^\pprime}
\left[{l^\prime(l^\prime+1)\over l^\pprime(l^\pprime+1)}\right]^{1/2}
\gamma^{m m^\prime m^\pprime}_{ll^\prime l^\pprime}
R_{l^\prime m^\prime}R_{l^\pprime m^\pprime}
-\sum_{l^\prime m^\prime}\sum_{l^\pprime m^\pprime}
[l^\prime(l^\prime+1)l^\pprime(l^\pprime+1)]^{1/2}
\gamma^{m m^\prime m^\pprime}_{ll^\prime l^\pprime}A_{l^\prime m^\prime}A_{l^\pprime m^\pprime}=0,
\label{expans_c}
\ee
where we have defined
\be
\gamma^{m m^\prime m^\pprime}_{l l^\prime l^\pprime}\equiv {1\over\sqrt{l^\prime(l^\prime+1)}}
\oint Y_{l^\prime m^\prime } {\bf B}_{l^\pprime m^\pprime}\cdot
{\bf C}^\ast_{lm}\;d\Omega. 
\ee
The expressions of the coupling coefficients $\gamma^{m m^\prime
m^\pprime}_{l l^\prime l^\pprime}$ in terms of the Wigner $3j$
symbols is given in Appendix~A. It is not difficult to see that, as
expected, this equation is trivially satisfied, since the two terms in
eq.~(\ref{expans_c}) are equal to the coefficients of the expansions
of $\nabla_\Omega R\times \nabla_\Omega V$ and $\nabla_\Omega A\times
\nabla_\Omega A$ in terms of ${\bf P}_{lm}$, both of which are zero.

We begin by expanding the vector product 
$\nabla_\Omega R\times \nabla_\Omega V$,
\be
\nabla_\Omega R\times \nabla_\Omega V=
\sum_{l^\prime m^\prime}\sum_{l^\pprime m^\pprime} 
R_{l^\prime m^\prime} V_{l^\pprime m^\pprime}
[l^\prime(l^\prime+1)l^\pprime(l^\pprime+1)]^{1/2}
({\bf B}_{l^\prime m^\prime}\times {\bf B}_{l^\pprime m^\pprime}).
\ee
Eliminating the coefficients $V_{l^\pprime m^\pprime}$ with eq.~(\ref{Vlm}), and using  
the relation (see Jones~1985),
\be
{\bf B}_{l^\prime m^\prime}\times {\bf B}_{l^\pprime m^\pprime}=-({\bf B}_{l^\prime m^\prime}\cdot
{\bf C}_{l^\pprime m^\pprime}){\bf\hat r},
\ee
we obtain
\be
\nabla_\Omega R\times \nabla_\Omega V=\sum_{l^\prime m^\prime}\sum_{l^\pprime m^\pprime}
\left[{l^\prime(l^\prime+1)\over l^\pprime(l^\pprime+1)}\right]^{1/2}
R_{l^\prime m^\prime}R_{l^\pprime m^\pprime}
({\bf B}_{l^\prime m^\prime}\cdot{\bf C}_{l^\pprime m^\pprime}){\bf\hat r}.
\ee
We now compute the expansion of this equation in series of ${\bf P}_{lm}$.
For this, we need to evaluate the coefficients
\[
\oint Y^\ast_{lm}{\bf B}_{l^\prime m^\prime}\cdot{\bf C}_{l^\pprime m^\pprime}\; d\Omega
=\oint [Y_{lm}{\bf B}^\ast_{l^\prime m^\prime}\cdot{\bf C}^\ast_{l^\pprime m^\pprime}]^\ast\; d\Omega
\]
\[
=(-1)^{m^\prime}\oint [Y_{lm}{\bf B}_{l^\prime-m^\prime}\cdot{\bf C}^\ast_{l^\pprime m^\pprime}]^\ast\; d\Omega
=(-1)^{m^\prime}\sqrt{l(l+1)}[\gamma^{m^\pprime m, -m^\prime}_{l^\pprime l l^\prime}]^\ast
\]
\be
=\sqrt{l(l+1)}[\gamma^{m m^\pprime m^\prime}_{l l^\pprime l^\prime}]^\ast
=-\sqrt{l(l+1)}[\gamma^{m m^\prime m^\pprime}_{l l^\prime l^\pprime}]^\ast
=\sqrt{l(l+1)}\gamma^{m m^\prime m^\pprime}_{l l^\prime l^\pprime},
\ee
where we have used the definition of $\gamma^{m m^\prime m^\pprime}_{l
l^\prime l^\pprime}$ and the symmetry properties of Wigner $3j$
symbols.  Thus, we finally obtain
\be
\nabla_\Omega R\times \nabla_\Omega V=\sum_{l m}\sqrt{l(l+1)}{\bf P}_{lm}
\left\{
\sum_{l^\prime m^\prime}\sum_{l^\pprime m^\pprime}
\left[{l^\prime(l^\prime+1)\over l^\pprime(l^\pprime+1)}\right]^{1/2}
\gamma^{m m^\prime m^\pprime}_{l l^\prime l^\pprime}
R_{l^\prime m^\prime}R_{l^\pprime m^\pprime}
\right\}.
\label{term1}
\ee
The same procedure, applied to the cross product $\nabla_\Omega A\times
\nabla_\Omega A$, shows that
\be
\nabla_\Omega A\times \nabla_\Omega A=-\sum_{l m}\sqrt{l(l+1)}{\bf P}_{lm}
\left\{
\sum_{l^\prime m^\prime}\sum_{l^\pprime m^\pprime}
[l^\prime(l^\prime+1)l^\pprime(l^\pprime+1)]^{1/2}
\gamma^{m m^\prime m^\pprime}_{l l^\prime l^\pprime}
A_{l^\prime m^\prime}A_{l^\pprime m^\pprime}
\right\}.
\label{term2}
\ee

Since $\nabla_\Omega R\times \nabla_\Omega V=0$, as can be shown
by taking the curl of eq.~(\ref{tanfor}), and $\nabla_\Omega
A\times \nabla_\Omega A=0$, we conclude that for each $(l,m)$ the
quantities inside curly brakets in eq.~(\ref{term1}) and (\ref{term2})
must be zero. Thus, each term of the expansion of the equation of 
force balance along ${\bf C}_{lm}$, eq.~(\ref{expans_c}), is zero.

\clearpage

\begin{deluxetable}{lllll}
\footnotesize
\tablecaption{\sc Multipole Coefficients for $H_0=0$ and $m=0$}
\tablewidth{0pt}
\tablehead{
\colhead{$l_{\rm max}$} & \colhead{$R_1$} & \colhead{$R_2$} & \colhead{$R_3$}
& \colhead{$R_4$} }
\startdata
2     & 2 & 0.7847 &        &        \\
3     & 2 & 0.7631 & 0.2604 &        \\
4     & 2 & 0.7623 & 0.2546 & 0.0792 \\
\hline
exact & 2 & 0.7623 & 0.2542 & 0.0791 \\
\enddata
\end{deluxetable}

\begin{deluxetable}{llllllll}
\footnotesize
\tablecaption{\sc Multipole Coefficients for $H_0=1/2$ and $m=0$}
\tablewidth{0pt}
\tablehead{
\colhead{$l_{\rm max}$} 
& \colhead{$A_1$} & \colhead{$R_2$} 
& \colhead{$A_3$} & \colhead{$R_4$} 
& \colhead{$A_5$} & \colhead{$R_6$} 
& $\lambda_r$
}
\startdata
2     &  3.0670 & $-2.3780$ &           &        &        &           & 2.002 \\
3     &  3.0583 & $-3.2861$ & $-0.2670$ &        &        &           & 1.943 \\
4     &  3.0591 & $-3.2573$ & $-0.2581$ & 0.6839 &        &           & 1.944 \\
5     &  3.0593 & $-3.2496$ & $-0.2532$ & 0.8388 & 0.0356 &           & 1.940 \\
6     &  3.0593 & $-3.2496$ & $-0.2534$ & 0.8249 & 0.0324 & $-0.1523$ & 1.941 \\
\enddata
\end{deluxetable}

\clearpage

\begin{figure}
\plotone{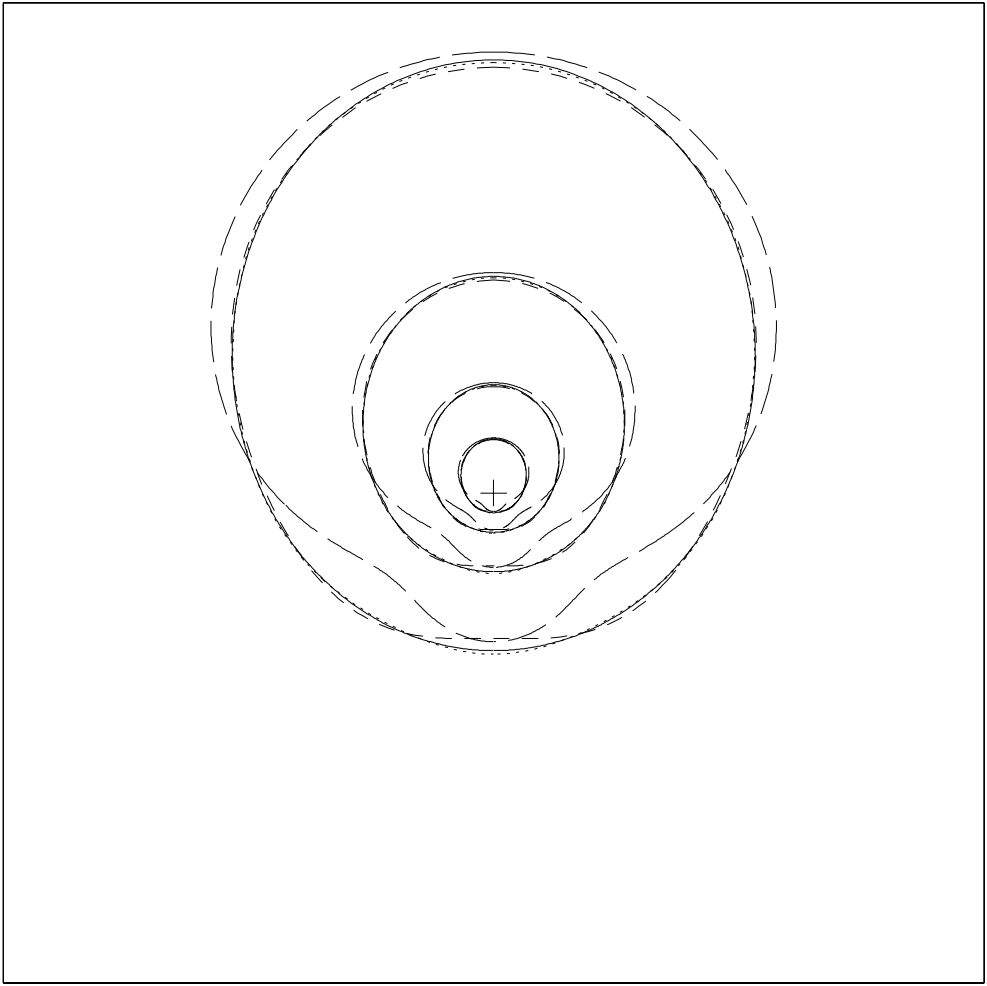}
\caption{Iso-density contours for the axisymmetric hydrostatic ($H_0=0$)
equilibrium with $R_1=2$ obtained with the method described in this paper
({\it long-dashed curves}, $l_{\rm max}=2$; {\it short-dashed curves},
$l_{\rm max}=3$; {\it dotted curves} $l_{\rm max}=4$).  The analytical
solution of MN00 with the same value of $R_1$ is shown by the {\it
solid curves}.}
\label{h0m0}
\end{figure}

\clearpage

\begin{figure}
\plotone{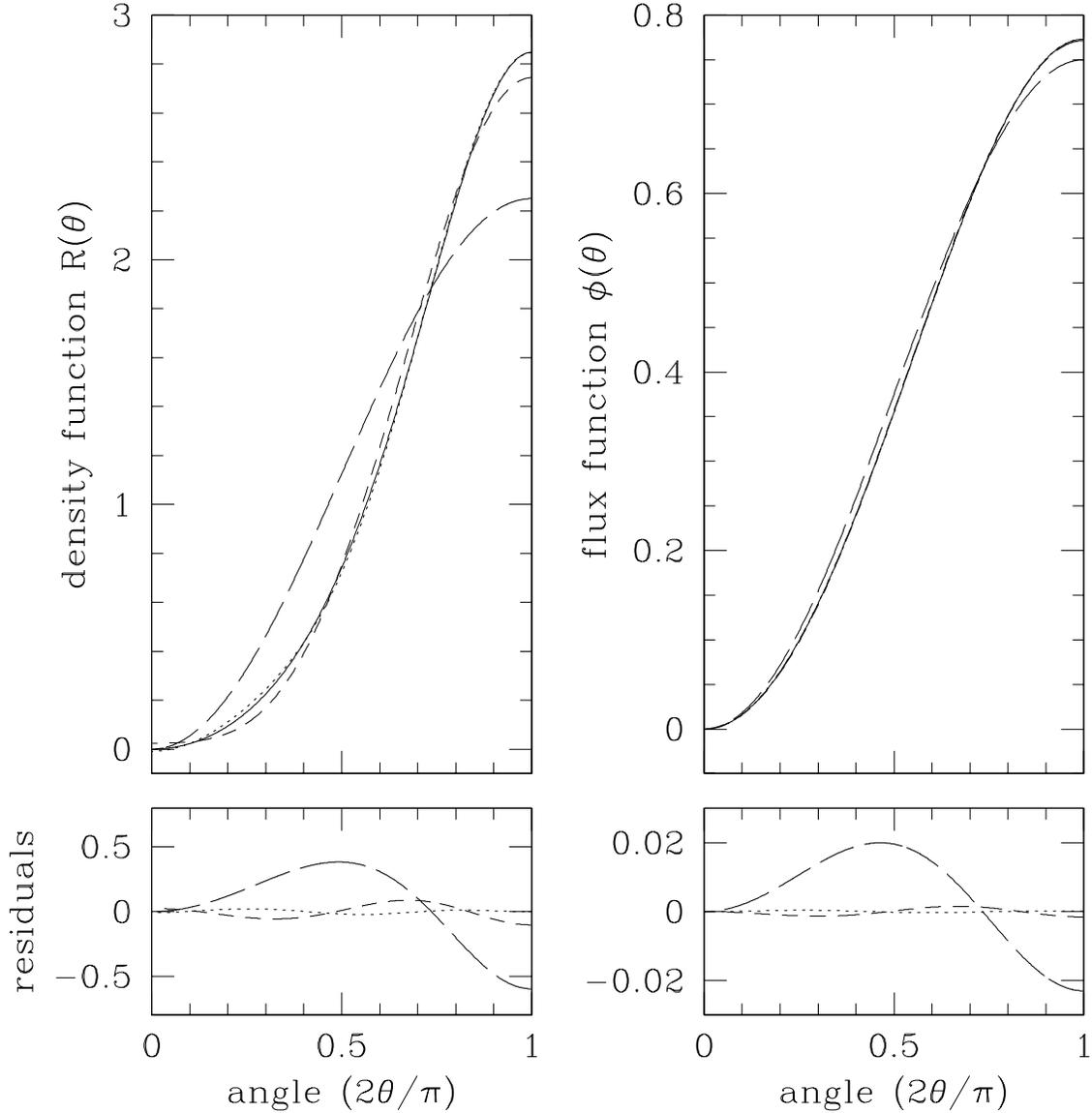}
\caption{Density function $R(\theta)$ and flux function $\phi(\theta)$
for the axisymmetric case with $H_0=0.5$. The {\it solid curves} are
the numerical solutions of LS96. The solutions obtained with the method
described in this paper for $m=0$ are shown by {\it long-dashed curves}
($l_{\rm max}=2$), {\it short-dashed curves} ($l_{\rm max}=4$), and {\it
dotted curves} ($l_{\rm max}=6$). The lower panels show the differences
between the exact solution of LS96 and the series solution 
obtained in this work.}
\label{h05m0_a}
\end{figure}

\clearpage

\begin{figure}
\plotone{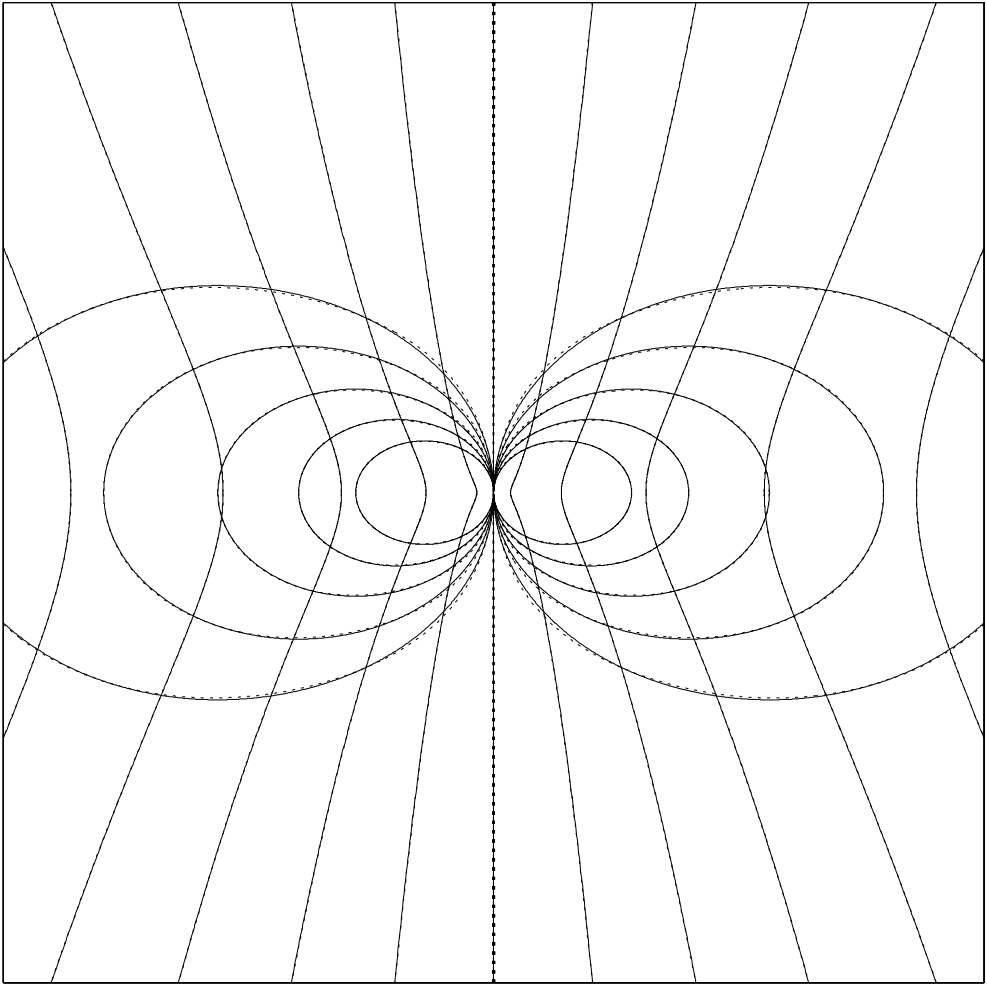}
\caption{Axisymmetric magnetostatic equilibrium for $H_0=0.5$ 
(isodensity contours and magnetic field lines).
The solution of LS96 is shown by {\it solid lines}, the solution
obtained in this paper with $l_{\rm max}=6$ is shown by {\it dotted lines}.
The agreement of the two solutions is very good.}
\label{h05m0_b}
\end{figure}

\clearpage

\begin{figure}
\plotone{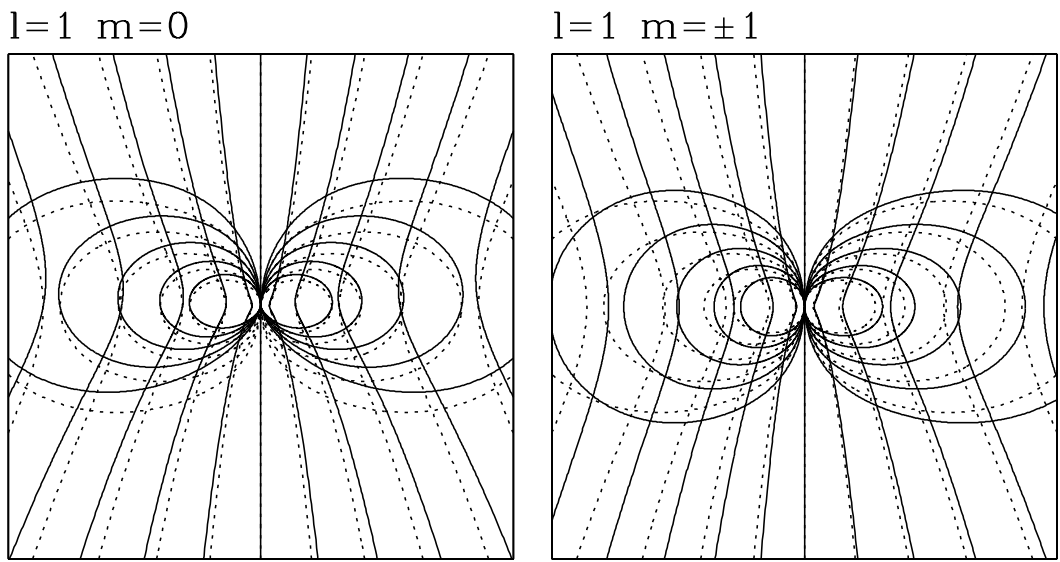}
\caption{The $l=1$ gauge. For any value of $H_0$, magnetostatic equilibria
with azimuthal and equatorial symmetry ({\it dashed lines}) possess
neighboring equilibria with a non-zero $l=1$ density component ({\it
solid lines}). The two panels show examples of these equilibria with small
density perturbations containing $l=1,m=0$ and $l=1,m=\pm 1$ harmonics.}
\label{l1}
\end{figure}

\clearpage

\begin{figure}
\plotone{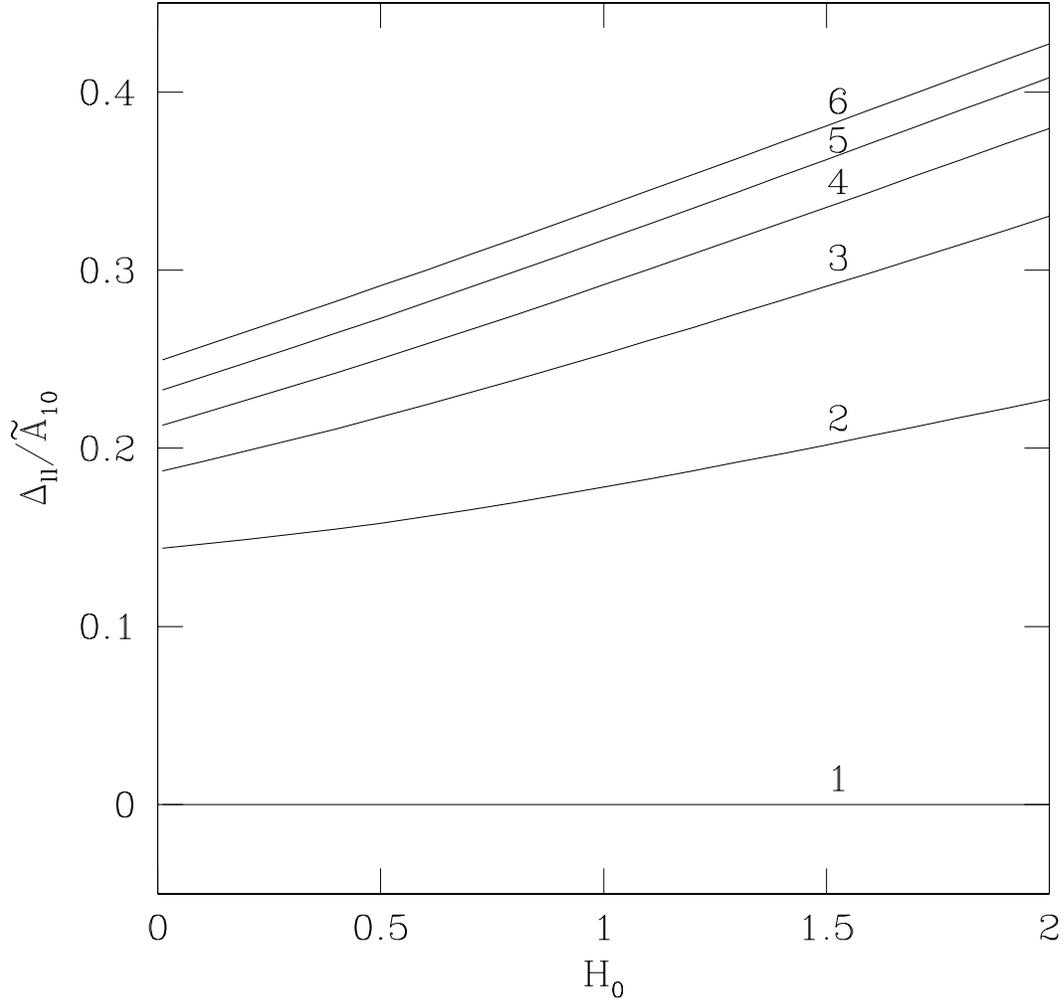}
\caption{The determinant $\Delta_{ll}$ of the system of linearized
equations describing sectorial ($l=m$) distortions of 
axisymmetric magnetostatic equilibria. The determinant is normalized
to $\tilde A_{10}$, the value of the dipole term in the expansion of
the vector potential for the magnetic field. Notice that the determinant
is zero for any value of $H_0$ for $l=1,m=\pm 1$ density distortions,
as discussed in Sect.~7.2, and non-zero in all other cases.}
\label{det}
\end{figure}

{}

\end{document}